\documentclass[twoside]{LCWS11}
\usepackage[latin1]{inputenc}
\usepackage[dvips]{graphicx,epsfig,color}
\usepackage{wrapfig,rotating}
\usepackage{amssymb,amsmath,array}
\begin{document}
\title{
Feasibility study of measurement of Higgs pair creation in a $\gamma \gamma$ collider} 
\author{Shin-ichi Kawada$^1$, Katsumasa Ikematsu$^2$, Tohru Takahashi$^1$,\\
Keisuke Fujii$^3$, Yoshimasa Kurihara$^3$
\vspace{.3cm}\\
1- Advanced Sciences of Matter, Hiroshima University \\
1-3-1, Kagamiyama, Higashi-Hiroshima, Hiroshima, 739-8530, Japan
\vspace{.1cm}\\
2- Department f\"{u}r Physik, Universt\"{a}t Siegen \\
D-57068, Siegen, Germany
\vspace{.1cm}\\
3- High Energy Accelerator Research Organization (KEK) \\
1-1, Oho, Tsukuba, Ibaraki, 305-0801, Japan
}

\maketitle

\begin{abstract}
We studied the feasibility of measurement of Higgs pair creation in a $\gamma \gamma$ collider.
We found the optimum collision energy is around 270 GeV from the sensitivity study with Higgs boson mass of 120 GeV/$c^2$.
Main backgrounds are $\gamma \gamma \rightarrow WW$, $\gamma \gamma \rightarrow ZZ$, and $\gamma \gamma \rightarrow b\bar{b}b\bar{b}$ at the optimum collision energy.
The preliminary analysis shows $\gamma \gamma \rightarrow HH$ could be observed with the statistical significance $\sim 5 \sigma$ when we chose correct assignment of a track by using color singlet information.
\end{abstract}

\section{Introduction}

\begin{wrapfigure}{r}{200pt}
\vspace*{-10pt}
\centering
\includegraphics[width = 6cm, clip]{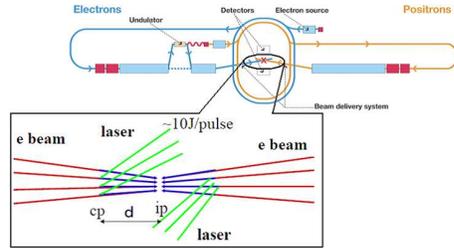}
\vspace*{-10pt}
\caption{An outline of PLC.
Positron beam replaced to electron beam.}
\label{fig:PLC}
\end{wrapfigure}

A Photon Linear Collider (PLC) has been considered as a possible option of the International Linear Collider (ILC).
In the PLC, high energy photons are generated by using inverse Compton scattering between electrons and laser photons.
An outline of PLC is shown in Figure \ref{fig:PLC}.
Details of PLC are found in \cite{PLC}.

\begin{figure}[!h]
\centering
\includegraphics[scale = 0.35, clip]{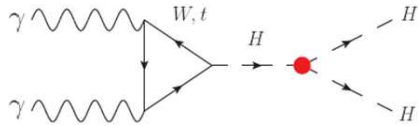}
\caption{A diagram of Higgs pair production by $\gamma \gamma$ collision via the self-coupling.}
\label{fig:AAHH}
\end{figure}

In this study, we investigated the feasibility of studying the self-coupling of Higgs boson.
The Higgs self-coupling constant $\lambda$ can be represented as $\lambda = \lambda ^{\mathrm{SM} }(1 + \delta \kappa)$ which contributes via the diagram shown in Figure \ref{fig:AAHH} to the Higgs pair production in photon-photon collision.
$\lambda ^{\mathrm{SM}}$ is the Higgs boson self-coupling constant which is included in the Standard Model, and $\delta \kappa$ is the deviation from the Standard Model.

Theoretical studies have been performed for this process by several authors~\cite{HSC, Asakawa:2008se, Asakawa:2009ux}.
However, the cross-section of Higgs self-coulping is very small, experimental feasibility against huge backgrounds is yet to be studied.
In this study, we investigated feasibility to observe Higgs pair production in high energy photon collision by Monte-Calro simulation using a realistic luminosity spectrum based on a set of PLC parameters.

\section{Optimization of $\gamma \gamma$ collision energy}

In order to decide the optimum collision energy, we defined the statistical sensitivity for the $\delta \kappa$ as;
\begin{equation*}
\mathrm{sensitivity} = \dfrac{\left| N(\delta \kappa) - N_{\mathrm{SM}} \right|}{\sqrt{N_{\mathrm{obs}}}} = \dfrac{L\left| \eta \sigma (\delta \kappa) - \eta \sigma _{\mathrm{SM}} \right|}{\sqrt{L(\eta \sigma (\delta \kappa) + \eta _{\mathrm{BG}}\sigma _{\mathrm{BG}})}}
\end{equation*}
where, $N(\delta \kappa)$ is the expected number of events as a function of $\delta \kappa$, and $N_{\mathrm{SM}}$ is the expected number of events from the Standard Model.
$\sigma (\delta \kappa)$ and $\sigma _{\mathrm{SM}}$ are the cross-section of Higgs boson production as a function of $\delta \kappa$ and for the Standard Model.
$L$, $\eta$, $\eta _{\mathrm{BG}}$, and $\sigma _{\mathrm{BG}}$ are the integrated luminosity, detection efficiency for signal, detection efficiency for background, and the cross-section of background processes, respectively.
For $\eta = 1$ and $\eta _{\mathrm{BG}} = 0$, the sensitivity is written as;
\begin{equation*}
\mathrm{sensitivity} = \sqrt{L} \dfrac{\left| \sigma (\delta \kappa) - \sigma _{\mathrm{SM}} \right|}{\sqrt{\sigma (\delta \kappa)}}.
\end{equation*}

\begin{wrapfigure}{r}{160pt}
\vspace*{-10pt}
\centering
\includegraphics[width = 5.5cm, clip]{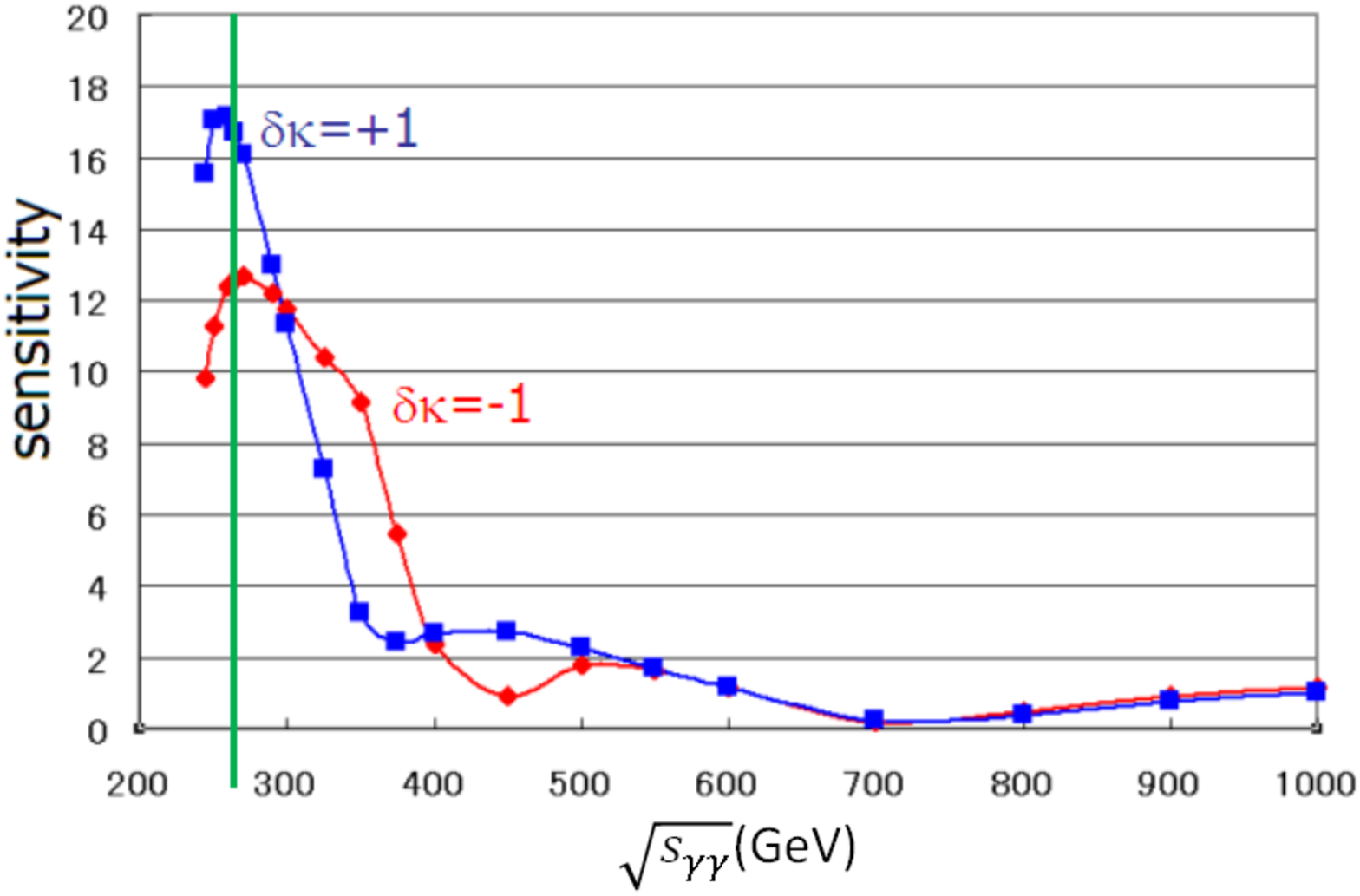}
\caption{Sensitivity plot for the function of $\sqrt{s_{\gamma \gamma}}$.
Green line shows the optimum energy, 270 GeV.}
\label{fig:sensitivity}
\vspace*{-20pt}
\end{wrapfigure}

Figure \ref{fig:sensitivity} shows the sensitivity plot as a function of the energy of $\gamma \gamma$ interaction, $\sqrt{s_{\gamma \gamma}}$, for the Higgs boson mass of 120 GeV/$c^2$ with integrated luminosity of 1000 fb$^{-1}$.
The cross-section of signal was calculated by using the formula given in \cite{Asakawa:2008se} for the case of $\delta \kappa = \pm 1$ as indicated in Figure \ref{fig:sensitivity}.
From this result, we set $\sqrt{s_{\gamma \gamma}} = 270$ GeV for this study. 

\section{Beam parameters}

The parameters for the electron and laser beams are chosen to maximize $\gamma \gamma$ luminosity around $\sqrt{s_{\gamma \gamma}} = 270$ GeV based on TESLA optimistic parameters~\cite{TESLA}, where we assumed the same parameters with the TESLA except for the electron beam energy (Table \ref{tab:param}).
The luminosity distribution for the parameters was simulated by CAIN~\cite{CAIN} and is shown in Figure \ref{fig:lumi}.

\begin{table}[!h]
\begin{tabular}{cc}
\begin{minipage}{0.5\hsize}
\centering
\begin{tabular}{cc} \hline
$E_e$ [GeV] & 190 \\
$N/10^{10}$ & 2 \\
$\sigma _z$ [mm] & 0.35 \\
$\gamma \varepsilon _{x/y}$ [10$^{-6}$mrad] & 2.5/0.03 \\
$\beta _{x/y}$ [mm] \verb|@| IP & 1.5/0.3 \\
$\sigma _{x/y}$ [nm] & 96/4.7 \\
$\lambda _L$ [nm] & 1054 \\
Pulse energy [J] & 10 \\
$x = 4\omega E_e/m^2_e$ & 3.76 \\ \hline
\end{tabular}
\caption{The parameters of electron and laser beams.}
\label{tab:param}
\end{minipage}
\begin{minipage}{0.5\hsize}
\centering
\makeatletter
\def\@captype{figure}
\makeatother
\includegraphics[width = 5cm, clip]{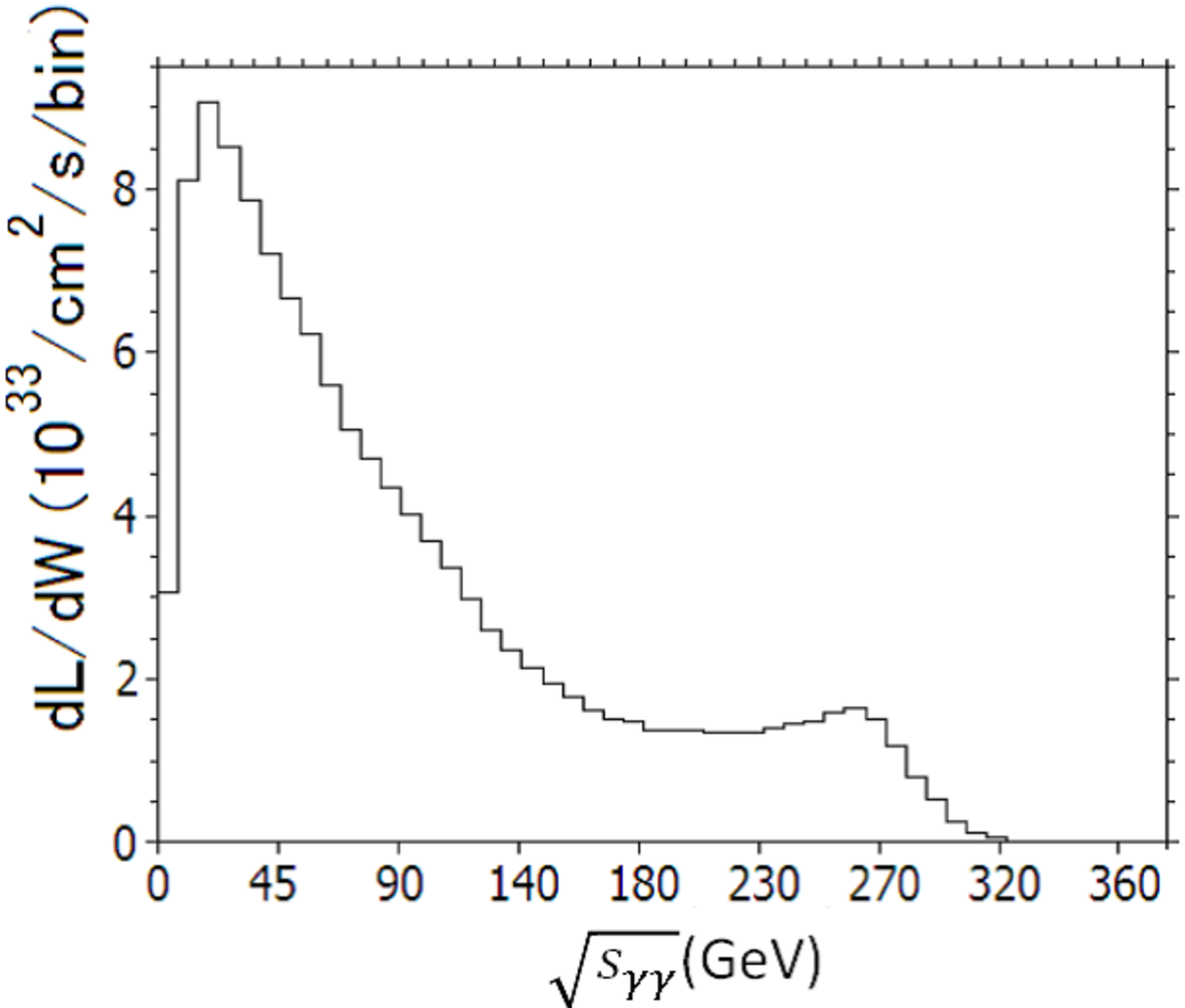}
\vspace*{-10pt}
\caption{Luminosity distribution simulated by CAIN.}
\label{fig:lumi}
\end{minipage}
\end{tabular}
\end{table}

\section{Signal and backgrounds}

\begin{figure}[!h]
\centering
\includegraphics[scale = 0.37, clip]{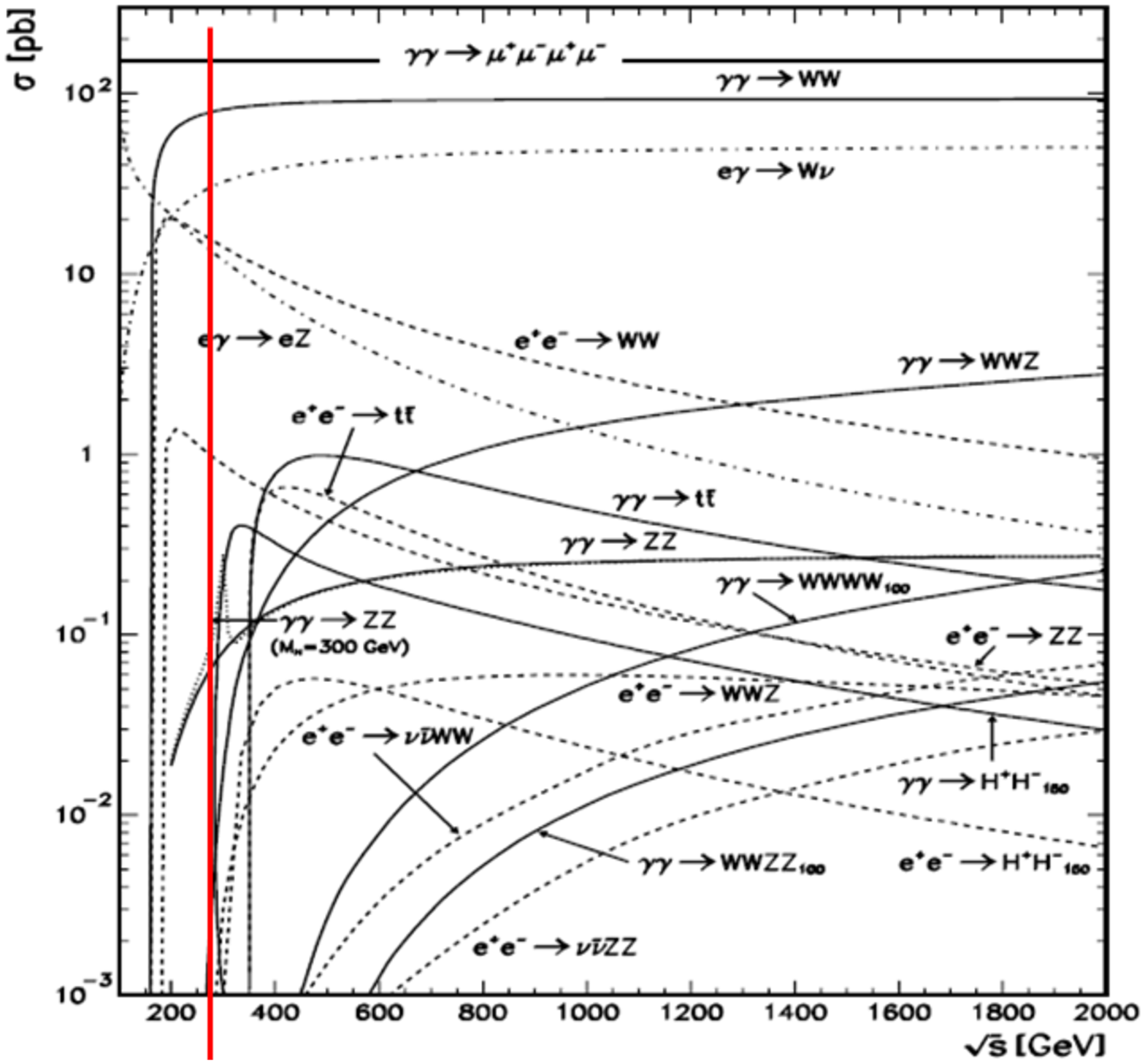}
\caption{Cross-sections of processes as a function of center mass energy.
The red line shows the optimum collision energy, 270 GeV.}
\label{fig:process}
\end{figure}

Figure \ref{fig:process} shows the cross-section for various $\gamma \gamma$ and $e^+e^-$ processes as a function of the center mass energy.
It indicates that the $\gamma \gamma \rightarrow WW$ and $\gamma \gamma \rightarrow ZZ$ processes will be the main backgrounds at $\sqrt{s_{\gamma \gamma}} = $ 270 GeV.

Since the main decay mode of the Higgs boson of 120 GeV/$c^2$ is $H \rightarrow b\bar{b}$ (branching ratio $\sim$ 0.68)~\cite{HiggsBR}, we consentrated on the case of $\gamma \gamma \rightarrow HH \rightarrow b\bar{b}b\bar{b}$.
Therefore, the $\gamma \gamma \rightarrow b\bar{b}b\bar{b}$ process is also considered as a background source.

Then we estimated the number of events by convoluting the luminosity distribution, {\it{i.e.}}:
\begin{equation*}
N_{\mathrm{events}} = \int \sigma (\sqrt{s_{\gamma \gamma}}) \dfrac{dL}{d\sqrt{s_{\gamma \gamma}}}d\sqrt{s_{\gamma \gamma}}
\end{equation*}
We caluculated $\gamma \gamma \rightarrow HH$ by using the formula in \cite{Asakawa:2008se,Asakawa:2009ux}, $\gamma \gamma \rightarrow WW$ by using HELAS~\cite{HELAS}, $\gamma \gamma \rightarrow ZZ$ by using gamgamZZ-code~\cite{ZZ-1,ZZ-2}, and $\gamma \gamma \rightarrow b\bar{b}b\bar{b}$ by using GRACE~\cite{GRACE}.
The numerical integration and event generation was performed by a Monte-Calro integration and event generation program BASES/SPRING~\cite{Kawabata:1995th}.
We expect 16 events/year for $\gamma \gamma \rightarrow HH$, $1.462 \times 10^7$ events/year for $\gamma \gamma \rightarrow WW$, $1.187 \times 10^4$ events/year for $\gamma \gamma \rightarrow ZZ$.
For $\gamma \gamma \rightarrow b\bar{b}b\bar{b}$, $5.872 \times 10^4$ events/year is expected for events with $b\bar{b}$ mass of grater than 15 GeV/$c^2$.

\section{Simulation and analysis}

JSF (JLC Study Framework)~\cite{JSF1,JSF2} was used as the simulation framework in this study.
Pythia~\cite{Pythia} was used for the parton shower and hadronization.
A fast simulator~\cite{JSF2} was used for the detector simulation instead of full simulation.
We generated $5 \times 10^4$ Monte-Calro events for $\gamma \gamma \rightarrow HH$, $7.5 \times 10^7$ for $\gamma \gamma \rightarrow WW$, $1 \times 10^6$ for $\gamma \gamma \rightarrow ZZ$, and $1 \times 10^6$ for $\gamma \gamma \rightarrow b\bar{b}b\bar{b}$.

First, we applied the forced 4-jet clustering to reconstruct the jets by using JADE clustering algorithm~\cite{JADE}.
In order to select proper combination of reconstructed jets, $\chi ^2_i$s ($i = H, W, Z, b\bar{b}$) were calculated as;
\begin{equation*}
\chi ^2_i = \mathrm{min}\left[ \dfrac{\left( M_1 - M_i \right) ^2}{\sigma _{2ji}} + \dfrac{\left( M_2 - M_i \right) ^2}{\sigma _{2ji}} \right],
\end{equation*}
where $M_1$ and $M_2$ are reconstructed particle mass.
$M_i \> (i = H, W, Z, b\bar{b})$ are the mass of Higgs boson, $W$ boson, $Z$ boson, and the invariant mass of $b\bar{b}$ (10 GeV).
$\sigma_{2ji} \> (i = H, W, Z, b\bar{b})$ are the mass resolutions for $H$, $W$, $Z$, and $b\bar{b}$, respectively.
The jet combination which has the least $\chi ^2_i$ was considered as the most probable combination.

\begin{wrapfigure}{r}{170pt}
\centering
\vspace*{-10pt}
\includegraphics[width = 5.5cm, clip]{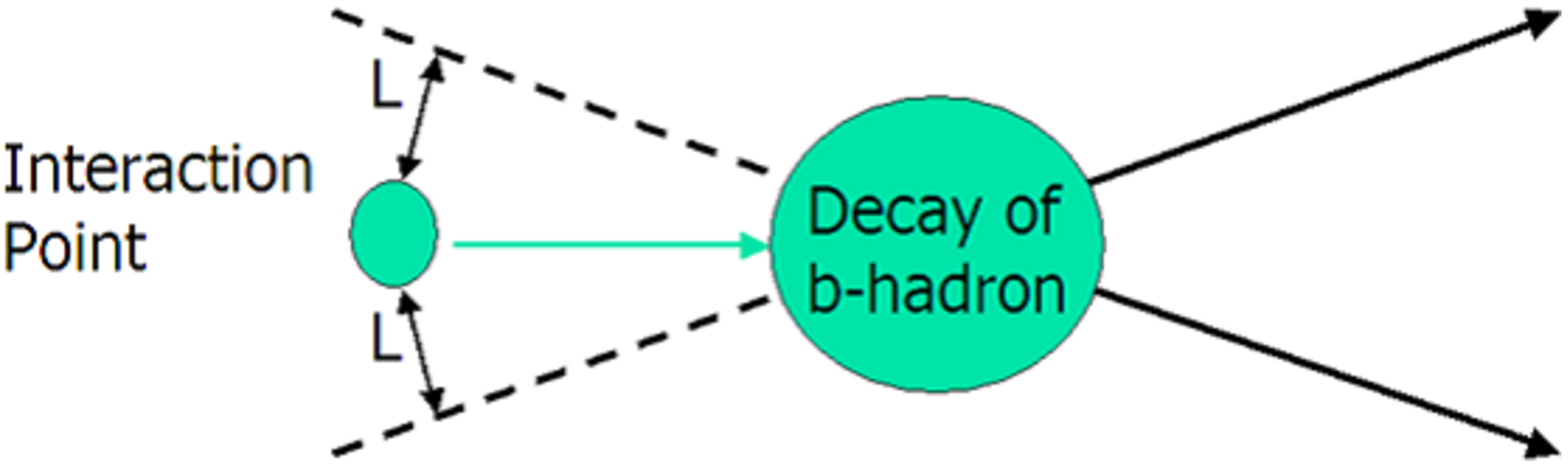}
\caption{An outline of "nsig" method.
$b$-hadron is generated at "Interaction Point" and decay at "Decay of $b$-hadron".
Arrows and dotted lines are the particle tracks and extrapolate particle tracks.}
\label{fig:nsig}
\end{wrapfigure}

In order to distinguish $b$ quarks, we used "nsig" method in this study.
Figure \ref{fig:nsig} shows the outline of "nsig" method.
For each charged track in a reconstruct jet, $N_{\mathrm{sig}} \equiv L/\sigma _L$ was calculated, where $L$ is the closest approach to the "Interaction Point" of the track in the plane perpendicular to the beam, and $\sigma _L$ is its resolution.
Then $N_{\mathrm{offv}}(a)$, the number of tracks which have $N_{\mathrm{sig}} > a$, is calculated for each jet.

Before optimizing selection criteria, we applied the pre-selection:
\begin{itemize}
\item $N_{\mathrm{jet}}(N_{\mathrm{offv}}(3.0) \ge 1) \ge 3$,
\item $N_{\mathrm{jet}}(N_{\mathrm{offv}}(3.0) \ge 2) \ge 2$,
\end{itemize}
where $N_{\mathrm{jet}}(N_{\mathrm{offv}}(3.0) \ge b)$ is the number of jets in which  $N_{\mathrm{offv}}(3.0)$ is greater than or equal to $b$.

Then we applied the Neural Network analysis to optimize the event selection criteria.
JETNET~\cite{NN} was used to train and to evaluate performance of the Neural Network.
The Neural Network was trained for each background separately so as to maximize the statistical significance $\Sigma$ defined by
\begin{equation*}
\Sigma \equiv \dfrac{N_{\mathrm{signal}}}{\sqrt{N_{\mathrm{signal}} + N_{\mathrm{BG}}}},
\end{equation*}
where $N_{\mathrm{signal}}$ and $N_{\mathrm{BG}}$ are the number of signal and background events.
Table \ref{tab:JADE} shows the cut statistics with JADE clustering.
From Table \ref{tab:JADE}, the $\Sigma$ was calculated to be $1.17 \sigma$.

\begin{table}[!h]
\centering
\begin{tabular}{ccccc} \hline
\rule[0pt]{0pt}{0pt} & & & & \\[-9pt]
 & $HH$ & $WW$ & $ZZ$ & $b\bar{b}b\bar{b}$ \\ \hline
\rule[0pt]{0pt}{0pt} & & & & \\[-9pt]
expected events & 80 & $7.31 \times 10^7$ & 59350 & 293600 \\
pre-selection & 47.93 & 81655 & 5167 & 84491 \\
$W$ filter & 12.34 & 8.772 & 193.4 & 568.4 \\
$b\bar{b}$ filter & 8.238 & 0 & 84.40 & 13.21 \\
$Z$ filter & 4.994 & 0 & 7.359 & 5.872 \\ \hline
\end{tabular}
\caption{Cut statistics with JADE clustering.}
\label{tab:JADE}
\end{table}

To investigate possible improvement of event selection and background suppression, we chose correct assignment of tracks to each jet as well as correct selection of jet pairs from a parent particle by using color singlet information from event generators.
It was applied to $\gamma \gamma \rightarrow HH, WW, ZZ$, but not to $\gamma \gamma \rightarrow b\bar{b}b\bar{b}$, because color singlet combination for $b\bar{b}$ pair in $b\bar{b}b\bar{b}$ event is not trivial.
Table \ref{tab:cheat} shows the cut statistics after optimizing the Neural Network for perfect jet clustering.
From Table \ref{tab:cheat}, the $\Sigma$ was calculated to be $4.95 \sigma$.
This result indicates that the $\gamma \gamma \rightarrow HH$ could be observed with $\sim 5 \sigma$ with the integrated luminosity corresponds to 5-year operation of the PLC, if we could improve performance of the jet clustering and $b$-tagging algorithm.

\begin{table}[!h]
\centering
\begin{tabular}{ccccc} \hline
\rule[0pt]{0pt}{0pt} & & & & \\[-9pt]
 & $HH$ & $WW$ & $ZZ$ & $b\bar{b}b\bar{b}$ \\ \hline
\rule[0pt]{0pt}{0pt} & & & & \\[-9pt]
expected events & 80 & $7.31 \times 10^7$ & 59350 & 293600 \\
pre-selection & 46.64 & 55836 & 4172 & 84491 \\
$W$ filter & 38.58 & 4.873 & 98.84 & 2179 \\
$b\bar{b}$ filter & 34.50 & 2.924 & 27.76 & 2.642 \\
$Z$ filter & 33.06 & 2.924 & 5.935 & 2.642 \\ \hline
\end{tabular}
\caption{Cut statistics with perfect clustering.}
\label{tab:cheat}
\end{table}

\section{Summary}

We studied the feasibility of measurement of Higgs pair creation in the PLC particulary against large background processes such as $\gamma \gamma \rightarrow WW$, $\gamma \gamma \rightarrow ZZ$, and $\gamma \gamma \rightarrow b\bar{b}b\bar{b}$.
The optimum collsion energy was found to be 270 GeV from the sensitivity study with Higgs boson mass of 120 GeV/$c^2$.
The result is preliminary in a sense that systematic error of the Neural Network analysis is yet to be evaluated, but it indicates that $\gamma \gamma \rightarrow HH$ could be observed with $\sim 5 \sigma$ with the integrated luminosity corresponds to 5-year operation of the PLC if we could choose correct assignment of tracks by using color singlet information.

\section{Acknowledgments}

The authors would like to thank ILC physics working group~\cite{ILCphys} for useful discussions.
This work is supported in part by the Creative Scientific Research Grant No. 18GS0202 of the Japan Society for Promotions of Science (JSPS), the JSPS Core University Program, and the JSPS Grant-in-Aid for Science Research No. 22244031, and the JSPS Specially Promoted Research No. 23000002.


\begin{footnotesize}


\end{footnotesize}


\end{document}